          [2001/04/25 1.1 (PWD)]
\documentclass[a4paper,twocolumn]{esapub} % European paper

\usepackage{amssymb}

\newcommand{\EE}{\mathcal{E}}

\newcommand{\ee}{{\sf e}}
\newcommand{\kk}{{\sf k}}

\def\lprox{\mathrel{\raise .3ex\hbox{$<$\kern-
.75em\lower1ex\hbox{$\sim$}}}}

\def\gprox{\mathrel{\raise .3ex\hbox{$>$\kern-
.75em\lower1ex\hbox{$\sim$}}}}

\title{Problems with wormholes which involve arbitrarily small amounts of exotic matter}
\author{C. J. Fewster}
\affil{Department of Mathematics, University of York, Heslington,
York YO10 5DD, United Kingdom, Email: cjf3@york.ac.uk}
\author{T. A. Roman}
\affil{Department of Mathematical Sciences, 
Central Connecticut State University, 
New Britain, CT 06050, Email: roman@ccsu.edu}

\begin{document}


\maketitle

\begin{abstract}
A quantum inequality bound on the expectation value of the null-contracted 
stress tensor in an arbitrary Hadamard state is used to obtain constraints 
on the geometries of traversable wormholes. Particular attention is given to the 
wormhole models of Visser, Kar, and Dadhich (VKD) and to those of Kuhfittig. These are 
models which use arbitrarily small amounts of exotic matter for wormhole 
maintenance. It is shown that macroscopic VKD models are either ruled out or severely 
constrained, while a recent model of Kuhfittig is shown to be, in fact, {\it non}-traversable.
\end{abstract}

\section{Introduction}
 
It has been known for some time that quantum field theory (QFT) allows violations 
of all known pointwise energy conditions (Epstein et~al., 1965) 
such as the ``weak energy condition'' (WEC):
\begin{equation}
T_{ab}\,u^{a}u^{b}  \geq 0 \,,
\label{eq:WEC}
\end{equation}
for all timelike vectors $u^{a}$, and the 
``null energy condition'' (NEC)
\begin{equation}
T_{ab}\,k^{a}k^{b} \geq 0 \,,
\label{eq:NEC}
\end{equation}
for all null vectors $k^a$. (For these pointwise conditions, 
the latter condition follows from the WEC by continuity.)
Some known examples of these violations are the Casimir effect 
[(Casimir, 1948), (Brown \& Maclay, 1969)], 
squeezed states (Slusher et~al, 1985), and the evaporation of black holes 
due to Hawking radiation [(Hawking, 1975), (Davies et~al, 1976), (Candelas, 1980)]. 
The first two have been experimentally observed (although the energy density 
itself is too small to be directly measurable). The third is theoretically predicted, but 
necessary to insure the consistency of black hole thermodynamics. 
All of these imply that we must take negative energy seriously.

On the other hand, if there are {\it no} constraints on negative energy, 
we could see gross macroscopic effects, such as 
\begin{itemize}
\item possible violations of 2nd law of thermodynamics (Ford, 1978)

\item traversable wormholes [(Morris and Thorne, 1988), (Visser, 1995)]

\item warp drive [(Alcubierre, 1994), (Krasnikov, 1998)]
 
\item time machines [(Morris et~al., 1988), (Visser, 1995), (Hawking, 1992)].
\end{itemize}

Over the last ten years, considerable progress has been made on 
the development of ``quantum inequality (QI)'' constraints on negative energy, 
which were first introduced by L. Ford (Ford, 1978, 1991). As a simple example, consider the massless scalar field 
in D-dimensional Minkowski spacetime ($\hbar=c=1$ units). It obeys a QI of the form: 
\begin{equation}
\int_{-\infty}^\infty \langle T_{ab}u^au^b\rangle_\omega
\,q(\tau)\,d\tau \ge -\frac{C}{\tau_0^D} \,,
\label{eq:QIform}
\end{equation}
where $q(\tau)$ is a smooth sampling function of width $\tau_0$, 
and normalized to have unit integral, and $C$ is a constant of order $1$ or less, 
which depends only on $q$ and $D$. The average is taken over a timelike geodesic, 
for all physically reasonable states $\omega$ and all sampling times
$\tau_0>0$. The class of sampling functions includes 
both (smooth) compactly and non-compactly supported sampling functions.

A variety of such bounds have been established over the last decade 
for massless and massive fields in flat and 
curved spacetimes for general worldlines and arbitrary Hadamard states. 
These include various extensions of the QI bounds to 
free Dirac, Maxwell, Proca, and Rarita--Schwinger fields. 
In addition, it has recently been proved that all unitary, positive 
energy conformal field theories in two-dimensional Minkowski space obey 
QI bounds, thus providing the first examples of QIs for interacting quantum field
theories. [For more details and references, see the review articles by Fewster (2003) and 
Roman (2004), as well as Fewster \& Roman (2005), which is a longer version 
of the present article.]

The QIs have been used to place strong constraints on exotic spacetimes, 
such as traversable wormhole and warp drive spacetimes 
[(Ford \& Roman, (1996), (Pfenning \& Ford, 1997), (Everett \& Roman, 1997), 
(Fewster, 2004), (Nandi et~al., 2004)]. 
The basic idea is to 
\begin{itemize}
\item 
obtain
the stress-energy tensor required to support a given spacetime from the metric 
and the Einstein equations; 
\item 
assume that this stress tensor is generated by quantum fields;
\item test it for consistency with the QI bounds, leading to
constraints on various parameters arising in the metric. 
\end{itemize}

The flat spacetime QI bounds have been applied to curved spacetimes under the 
following assumption. Let $\ell_{min}$ be the smallest 
proper radius of curvature (or smallest proper distance to any boundary in the spacetime). 
Spacetime is approximately Minkowskian in regions small compared to $\ell_{min}$, hence one 
can apply the flat spacetime QI-bounds in the ``short sampling time limit'': 
\begin{equation}
\tau_0=f \ell_{min} \,,
\end{equation} 
where $f\ll 1$ and the average is taken along timelike geodesics.

Strictly speaking, this is an {\em assumption}, but it is one for which 
good justification can be provided. The equivalence principle implies 
that physics ``in the small'' should be approximately Minkowskian as far 
as freely falling observers are concerned.  The assumption is also borne out 
by specific examples in four dimensions by taking the short-sampling time limit 
of various curved spacetime QIs [(Pfenning \& Ford, 1998), (Fewster \& Teo, 2000)]. 
Its validity has recently established for massless scalar fields in general
two-dimensional spacetimes (Fewster, 2004). 

\section{Morris-Thorne wormholes}
Morris and Thorne introduced a class of wormholes described by the metric:
\begin{equation}
ds^2=-e^{2\Phi(r(\ell))}dt^2 + {d\ell}^2 
                + {r^2}(\ell)({d\theta}^2+ {\rm sin}^2\theta\,{d\phi}^2) \,,      
\end{equation}
where $\ell$ is the radial proper length, and $\Phi$ is the ``red-shift function''. 
One can also write the metric in the form 
\begin{equation}
 ds^2=-e^{2\Phi(r)}dt^2+{{dr^2}\over{(1-b(r)/r)}}
           +r^2({d\theta}^2+ {\rm sin}^2\theta\,{d\phi}^2) \,,          
\end{equation}
where $b(r)$ is called the ``shape function''. The throat of the wormhole 
is located at $b(r)=r=r_0$ and 
\begin{equation}
\ell=\int_{r_0}^{\bar r(\ell)}{{dr}\over{(1-b(r)/r)^{1/2}}} \,,
\end{equation}
is the radial proper distance from $r_0$ to any $r>r_0$.

The curvature tensor components are 
\begin{equation}
R_{\hat{t}\hat{r}\hat{t}\hat{r}}
= \biggl(1- {b \over r} \biggr)\,[{\Phi}''\,+\,{({\Phi}')}^2]\, 
+\, { {\Phi}' \over {2r^2} } \, (b - b'r)  \,,     
\end{equation}
\begin{eqnarray}
R_{\hat{t}\hat{\theta}\hat{t}\hat{\theta}}
&=& R_{\hat{t}\hat{\phi}\hat{t}\hat{\phi}}
= { {\Phi}'\over r}\,\biggl(1- {b \over r} \biggr)  \,,  
                                            \label{eq:Rttheta} \\ 
R_{\hat{r}\hat{\theta}\hat{r}\hat{\theta}}  
&=& R_{\hat{r}\hat{\phi}\hat{r}\hat{\phi}}  
= {1 \over {2 r^3} } \, (b'r-b) \,,   \label{eq:Rrtheta}  \\
R_{\hat{\theta}\hat{\phi}\hat{\theta}\hat{\phi}}  
&=& b \over {r^3}  \,.             \label{eq:Rthetaphi}         
\end{eqnarray}
These are the components in a static orthonormal frame. 

The relevant stress tensor components for our discussion are 
\begin{equation}
\rho ={b' \over {8\pi r^2} } \,, 
\label{eq:Ttt} 
\end{equation}
\begin{equation}
p_r   =-{1 \over {8\pi} } \, \biggl[{ b \over {r^3} }-{ {2{\Phi}'} \over r }\,
 \biggl(1-{b \over r} \biggr) \biggr] \,. \label{eq:Trr} 
\end{equation}
The quantities $\rho$ and $p_r$ are the
mass-energy density and radial pressure, 
respectively, as measured by a static observer.  

Let $R_{max}$ be the magnitude of the maximum curvature component. 
Then the smallest proper radius of curvature is 
\begin{equation}
r_c \approx {1 \over {\sqrt {R_{max}}} }  \,.
\label{eq:l_c}
\end{equation} 
If we work in a small spacetime volume around the throat 
whose size is $\ll r_c$, then the flat QI bounds should then be applicable in 
this region. One typically finds (Ford \& Roman, 1996) that the QI bound implies that {\it either}  
the wormhole is only a little larger than Planck size {\it or} 
there is a large discrepancy in the length scales which characterize 
the wormhole, e.g., the (-) energy must be concentrated in a thin band 
{\it many orders of magnitude smaller} than the throat size. 
Similar constraints have been obtained for warp drives [(Pfenning \& Ford, 1997), (Everett \& Roman, 1997)].

\section{A New Approach} 

Most QIs to date have involved an average of the energy 
density over a timelike worldline. Recently a ``null-contracted'' QI   
was proven (Fewster \& Roman, 2003) in which one averages the {\it null}-contracted 
stress tensor $\langle T_{ab}\, k^a k^b\rangle_\omega$
over a {\it timelike worldline}. [The techniques used in the proof 
are those first used by Fewster (2000) to prove 
a general QI on energy density in arbitrary globally hyperbolic 
spacetimes.] As a simple case, consider the massless scalar 
field in 4D Minkowski spacetime where the average is taken over a timelike 
geodesic with (constant) four-velocity $u^a$, and where $k^a$ is a smooth constant 
null vector field. The new QI takes the form 
\begin{equation}
\int_{-\infty}^\infty d\tau \langle T_{ab}\, k^a k^b\rangle_\omega
g(\tau)^2  
\ge -\frac{(k_a u^a)^2}{12\pi^2} \int_{-\infty}^\infty d\tau\,g''(\tau)^2 \,,
\end{equation} 
for all Hadamard states $\omega$ and any smooth compactly
supported $g$. As before, we can apply this QI to wormhole spacetimes 
in the short sampling time limit, $\tau_0=f \ell_{min}$, where 
$f\ll 1$.

Suppose ${\langle T_{ab}k^a k^b \rangle}_{\omega}<\EE$, 
for $0\le \tau \le\tau_0$. Then
\begin{eqnarray}
\EE\int g(\tau)^2\, d\tau &\ge& \int {\langle T_{ab}k^a k^b \rangle}_{\omega}\, g(\tau)^2\,d\tau  \,,\\
&\ge& -\frac{1}{12\pi^2}\int g''(\tau)^2\,d\tau \,,
\end{eqnarray}
so
\begin{equation}
\EE \ge -\frac{1}{12\pi^2}\inf \frac{\int g''(\tau)^2\,d\tau}{\int g(\tau)^2\,d\tau} \,,
\end{equation}
taking the infimum over smooth $g$ compactly supported in 
$(0,\tau_0)$. Solving the variational problem one gets 
\begin{equation}
\EE \ge -\frac{C}{\tau_0^4} \,,\qquad C\sim 5 \,,
\end{equation}
(in units where $\hbar=c=1$). 

To apply our QI to wormhole spacetimes let us choose 
$\kk=\ee_{\hat{t}} + \ee_{\hat{r}}$, and our worldline to be that of a 
static observer at the throat. Then for this trajectory, $k_a u^a =1$, and  
\begin{equation}
T_{ab}\, k^a k^b=const=\frac{b_0'-1}{8\pi r_0^2l_p^2}<0 \,,
\label{eq:Tkk}
\end{equation}
where $b_0' = b'(r_0)$, and $l_p$ is the Planck length. 

Assume that the stress-energy tensor is generated by a Hadamard 
state of a free scalar quantum field. Then our QI bound implies:
\begin{equation}
\frac{1-b_0'}{8\pi r_0^2l_p^2} \le \frac{C}{\tau_0^4} \,.
\label{eq:QI-null}
\end{equation}
For $\tau_0 = f \ell_{min}$ we get
\begin{equation}
\frac{\ell^2_{min}}{r_0} \sqrt{1-b'_0} \le \frac{\sqrt{8\pi C} \, l_p}{f^2}\,.
\label{eq:premaster}
\end{equation}

As a reasonably generous interpretation of $\tau_0\ll \ell_{min}$, 
let us choose $f= 0.01$. Since $\sqrt{8\pi C}\approx 10.3$, we have that  
\begin{equation}
\frac{\ell^2_{min}}{r_0} \sqrt{1-b'_0} \lesssim 10^5 \, l_p\,.
\label{eq:master}
\end{equation} 
We can use this bound to obtain strong constraints on wormholes, in general. 
Let us now look at some specific cases.

\section{Visser-Kar-Dadich wormholes}
Visser, Kar, and Dadich (VKD) introduced the expression:
\begin{equation}
\int [\rho + p_r]\, dV = 2\int_{r_0}^{\infty}[\rho + p_r]\,4\pi r^2 \,dr 
\label{eq:total}
\end{equation}
as a suitable measure of the ``amount of exotic matter'' required to maintain a traversable 
wormhole (Visser et~al., 2003). [Other possible volume measures were discussed by Nandi et~al. (2004).] 
VKD then considered a class of wormholes for which the ``averaged null energy condition ANEC'' 
line integral is finite and negative, but for which the volume integral above can be made as 
small as one likes. They concluded that one can construct traversable (in principle) wormholes 
using only arbitrarily small amounts of exotic matter.

All of the models discussed by VKD are ``spatially Schwarzschild'', that is, 
\begin{equation}
b(r) = 2m = r_0 \,,
\end{equation}
so the $t=const$ spatial slices are the same as those of Schwarzschild. 
Therefore $b'(r)=0$, which implies that $\rho=0$, i.e., the energy density 
is zero in the static observer's frame. To apply the original QIs derived for energy 
density one would have to go to a boosted frame in which the energy density is negative. 
(Such a frame can always be found when the WEC is violated.) The disadvantage is that in 
so doing, one might have to insure that the sampling time does not exceed the proper time that 
the observer remains in the region of exotic matter. This problem does not arise using the 
{\it null}-contracted QI, because the latter involves not only the energy density but also the radial 
pressure. Hence we can take our observers to be static and located at the throat.

As a specific example of QI constraints on 
VKD wormholes, we consider the ``piecewise $R=0$ (zero scalar curvature)'' wormhole. 
For this type of wormhole 
\begin{equation}
{\rm exp}[\Phi(r)]= \epsilon + \lambda \sqrt{1-2m/r} \,,
\,\,\,\hfil\break 
{\rm for} \,\,\,r \in \, (r_0=2m,a) \,, 
\end{equation}
and
\begin{equation} 
{\rm exp}[\Phi(r)]= \sqrt{1-2m/r} \,, 
\,
\,\,\,{\rm for} \,\,\,r \in \,(a,\infty) \,,
\end{equation}
where $a$ is essentially the coordinate thickness of the region of exotic matter. 
Continuity of the metric coefficients implies that 
\begin{equation}
\lambda = 1 - \frac{\epsilon}{\epsilon_s} \,,
\end{equation}
where $\epsilon_s = \sqrt{1-2m/a}$. In this case, 
$\ell_{min} = r_c = r_0$. Applying our QI bound gives  
\begin{equation}
\frac{\ell^2_{min}}{r_0} \sqrt{1-b'_0} \lesssim 10^5 \, l_p\,,
\end{equation} 
with $b'_0=0$, we have
\begin{equation}
r_0 \lesssim 10^5 \, l_p \,.
\end{equation} 
So macroscopic ``piecewise $R=0$'' wormholes are ruled out.

It should also be noted that quite apart from the QI constraints, macroscopic 
VKD wormholes would be subject to other problems. The smaller the amount of exotic 
matter in these wormholes, the closer they are to being vacuum Schwarzschild wormholes. 
This could lead to problems with traversal, since the smaller the amount of $(-)$ 
energy, the longer the traversal time, as measured by external clocks. Possibly 
more serious are stabilization problems. The smaller the total amount of $(-)$ energy, 
the more unstable the wormhole is to even very small amounts of infalling $(+)$ energy radiation 
(e.g., infalling CMB photons), due to their enormous blueshifts.

\section{Kuhfittig's Wormholes}
Kuhfittig introduced a series of wormhole models (Kuhfittig, 1999, 2002, 2003) 
which were also designed to minimize the total amount of exotic matter. 
In Fewster \& Roman (2005), we used the QI to put severe 
constraints on one such class. Here we consider a recent model which 
he claims to be traversable, uses an arbitrarily small amount of exotic matter, 
and is consistent with the QIs (Kuhfittig, 2003). In this model he writes the metric in the form
\begin{equation}
ds^2=-e^{2\gamma(r)}dt^2 + e^{2\alpha(r)}dr^2 
                + {r^2}({d\theta}^2+ {\rm sin}^2\theta\,{d\phi}^2) \,,      
\end{equation}
so that 
\begin{equation}
b(r) = r(1 - e^{-2\alpha(r)}) \,.
\end{equation}
He then assumes that 
\begin{itemize}
\item (1) $\,\,\,\,\,\,{\rm lim}_{r \rightarrow r_0^+} \alpha(r) = + \infty$, 
so that $b'_0 \sim 1 $. 
\item (2) $\,\,\,\,\,\,\gamma(r)$ is finite at $r=r_0$.  
\end{itemize}
Assumption (1) implies that the 
wormhole flares outward very slowly. Assumption (2) is made to avoid an 
event horizon at the throat $r_0$.

In particular, he chooses 
\begin{equation}
\alpha(r) = \frac{k}{(r-r_0)^n} \,,\,\,\,\,\,\, n \geq 1 \,,
\label{eq:alpha-gen}
\end{equation}
where here $k$ is a (positive) constant and 
\begin{equation}
\gamma(r) = -\frac{L}{(r-r_2)^n} \,,\,\,\,\,\,\, n \geq 1 \,,
\label{eq:gamma-gen}
\end{equation}
where $L$ is another positive constant, and $0<r_2<r_0$. 
The choice $n \geq 1$ is made to obtain $b'(r) \sim 1$ near $r=r_0$. 
The condition on $r_2$ is imposed to avoid an event horizon at the throat $r_0$. 

These choices for $\alpha(r)$ and $\gamma(r)$ lead to the following problems:
\begin{itemize}
\item radially ingoing light rays reach 
the throat only after an {\it infinite lapse of affine 
parameter};

\item radially infalling particles take an 
{\it infinite amount of proper time} to reach the throat. 
\end{itemize}
Therefore, contrary to his claims, Kuhfittig's wormhole is 
{\it not} traversable. This failure is related to the very ``slow-flaring'' 
property, and an apparent confusion between proper and coordinate distance. 

As this contribution was being finalised, Kuhfittig released a preprint
(Kuhfittig 2005) which modifies the form of $\alpha(r)$ and $\gamma(r)$
near the throat to
\begin{equation}
\alpha(r) = \ln \frac{K}{(r-r_0)^a}
\end{equation}
and
\begin{equation}
\gamma(r) = -\ln \frac{L}{(r-r_2)^b}
\end{equation}
for positive constants $K$ and $L$, with $1/2<a<b<1$ and $0<r_2<r_0$.
With these choices the wormhole is now traversable, and has $b_0'=1$,
thereby
evading the general analysis given above. However, it suffers
from a different problem: one may show that points at radius $r=2r_0-r_2$
lie in the NEC-violating region for all possible values of the parameters
above, and that the radial tidal constraint
$|R_{\hat{t}\hat{r}\hat{t}\hat{r}}|\le {(10^8 {\rm m})}^{-2}$ (Morris and Thorne
1988) implies, for this value of $r$, that
\begin{equation}
\frac{{(r_0-r_2)}^{a-1}}{K} < {(10^8 {\rm m})}^{-1}
\end{equation}
(ignoring factors of order 1). But computing the proper distance to this
radius, $2r_0-r_2$, from the throat yields
\begin{equation}
\ell  = \frac{K {(r_0-r_2)}^{(1-a)}}{(1-a)} \ge 10^8 {\rm m} \,!
\end{equation}
So this model, as it stands, involves an extensive region of
NEC-violation, if it is to
respect the radial tidal constraint necessary for human traversal.
While the model can doubtless be modified further, to cut off the
NEC violation before radius $2r_0-r_2$, such a modification would have
to satisfy the quantum inequalities and tidal constraints at all radii
$r>r_0$ in the NEC-violating region
if it is to be a reasonable model. Whether this is possible or not remains
to be seen.

\section{Summary of results}
In this work we employed a recently derived QI bound on the {\it null}-contracted stress tensor 
averaged over a {\it timelike} worldline to give a simplified analysis of general wormhole models, 
not just those with small quantities of exotic matter. In particular our results imply 
that {\it macroscopic} VKD wormholes are ruled out or severely constrained, i.e., there is a large 
discrepancy between throat size and curvature radius. We also derived constraints on the Kuhfittig models, 
and in fact showed that one of the recent models presented is {\it not} traversable. 

The concentration of exotic matter to an arbitrarily small region around the wormhole throat, 
in a purely classical analysis, is by itself, not sufficient to guarantee both traversability 
and consistency with the QI bounds. 

\section*{Acknowledgments} 
The authors would like to thank Matt Visser, Larry Ford, 
and Lutz Osterbrink for helpful comments. 
This work was supported in part by the National
Science Foundation under Grant PHY-0139969 (to TAR). One of us (TAR) would like 
to thank the Mathematical Physics Group at the University of York for their kind 
hospitality during the course of some of this work.


\begin{thebibliography}{ZZ}


\bibitem{A94} Alcubierre M., 1994, Class. Quantum Grav. 11, L73  


\bibitem{BM69} Brown L.S., Maclay G.J., 1969, Phys. Rev. 184, 1272

\bibitem{C48} Casimir H.B.G., 1948, Proc. K. Ned. Akad. Wet. B51, 793  

\bibitem{C80} Candelas P., 1980, Phys. Rev. D 21, 2185

\bibitem{DFU76} Davies P.C.W., Fulling S.A., Unruh W.G., 1976, Phys. 
Rev. D 13, 2720   

\bibitem{EGJ65} Epstein H., Glaser V., Jaffe A., 1965, Nuovo Cim. 36, 1016 

\bibitem{ER97} Everett A.E., Roman T.A., 1997, Phys. Rev. D 56, 2100 

\bibitem{F00} Fewster C.J., 2000, Class. Quantum Grav. 17, 1897 
 

\bibitem{F03} Fewster C.J., Energy Inequalities in Quantum Field Theory, 
to appear in the Proceedings of the XIV ICMP, Lisbon 2003, math-ph/0501073 

\bibitem{F04} Fewster C.J., 2004, Phys. Rev. D 70, 127501 

\bibitem{FT00} Fewster C.J., Teo E., 2000, Phys. Rev. D 61, 084012 


\bibitem{FR03} Fewster C.J., Roman T.A., 2003, Phys. Rev. D 67, 044003 

\bibitem{FR05} Fewster C.J., Roman T.A., 2005, Phys. Rev. D 72, 044023 


\bibitem{F78} Ford L.H., 1978, Proc. Roy. Soc. Lond. A 364, 227
               

\bibitem{F91} Ford L.H., 1991, Phys. Rev. D 43, 3972 


\bibitem{FR96} Ford L.H., Roman T.A., 1996, Phys. Rev. D 53, 5496 


\bibitem{H75} Hawking S.W., 1975, Commun. Math. Phys. 43, 199  

\bibitem{H92} Hawking S.W., 1992, Phys. Rev. D 46, 603 

\bibitem{K98} Krasnikov S.V., 1998, Phys. Rev. D 57, 4760 


\bibitem{Ku99} Kuhfittig P.K.F., 1999, Am. J. Phys. 67, 125  

\bibitem{Ku02} Kuhfittig P.K.F., 2002, Phys. Rev. D 66, 024015 

\bibitem{Ku03} Kuhfittig P.K.F., 2003, Phys. Rev. D 68, 067502 

\bibitem{Ku05} Kuhfittig P.K.F., 2005, gr-qc/0508060


\bibitem{MT88} Morris M., Thorne K., 1988, Am. J. Phys. 56, 395    
   
\bibitem{MTU88} Morris M., Thorne K., Yurtsever U., 1988, Phys. Rev. Lett. 61, 
1446 

\bibitem{NZK04a} Nandi K.K, Zhang Y-Z, Kumar K.B.V., 2004, Phys. Rev. D 70, 064018 

\bibitem{NZK04b} Nandi K.K., Zhang Y-Z, Kumar K.B.V., 2004, Phys. Rev. D 70, 127503


\bibitem{PF97} Pfenning M.J., Ford L.H., 1997, Class. Quantum Grav. 14, 1743 

\bibitem{PF98} Pfenning M.J., Ford L.H., 1998, Phys. Rev. D 57, 3489 
 


\bibitem{SHYMV85} Slusher R.E., Hollberg L.W., Yurke B., Mertz J.C., and 
Valley J.F., 1985, Phys. Rev. Lett. 55, 2409  

  
\bibitem{R04} Roman T.A., Some Thoughts on Energy Conditions and Wormholes, 
to appear in the Proceedings of the 10th Marcel Grossmann Meeting on 
General Relativity and Gravitation, Rio de Janeiro, Brazil (2004), gr-qc/0409090



\bibitem{V95} Visser M., 1995, Lorentzian Wormholes - from Einstein to
Hawking, American Institute of Physics Press 



\bibitem{VKD03} Visser M., Kar S., Dadhich N., 2003, Phys. Rev. Lett. 90, 201102 



\end{thebibliography}
\end{document}